\documentclass[sigconf]{acmart}
\copyrightyear{2024} 
\acmYear{2024}
\setcopyright{acmlicensed}
\acmConference[MSR '24]{21st International Conference on Mining Software Repositories}{April 15--16, 2024}{Lisbon, Portugal}
\acmBooktitle{21st International Conference on Mining Software Repositories (MSR '24), April 15--16, 2024, Lisbon, Portugal}
\acmDOI{10.1145/3643991.3645074}
\acmISBN{979-8-4007-0587-8/24/04}

\usepackage{graphicx} 
\usepackage{booktabs}
\usepackage{tikz,lipsum,lmodern}
\usepackage{tcolorbox}
\usepackage{hyperref}
\usepackage{listings}
\usepackage{enumitem}
\usepackage{xcolor}
\usepackage{balance}

\newcommand{\mybox}[1]{%
	\setbox0=\hbox{#1}%
	\setlength{\@tempdima}{\dimexpr\wd0+13pt}%
	\begin{tcolorbox}[boxrule=0.5pt, colback=gray!10, arc=4pt,
		left=6pt,right=6pt,top=6pt,bottom=6pt,boxsep=0pt]
		#1
	\end{tcolorbox}
}

\author{Kailun~Jin,~Chung-Yu~Wang,~Hung~Viet~Pham,~Hadi~Hemmati}
\email{{ktaming,cywang14,hvpham,hemmati}@yorku.ca}
\affiliation{\institution{York University}\country{Toronto, ON Canada}}

\title{Can ChatGPT Support Developers? An Empirical Evaluation of Large Language Models for Code Generation}

\date{December 2023}

\begin{abstract}  

Large language models (LLMs) have demonstrated notable proficiency in code generation, with numerous prior studies showing their promising capabilities in various development scenarios. However, these studies mainly provide evaluations in research settings, which leaves a significant gap in understanding how effectively LLMs can support developers in real-world. To address this, we conducted an empirical analysis of conversations in DevGPT, a dataset collected from developers' conversations with ChatGPT (captured with the Share Link feature on platforms such as GitHub). Our empirical findings indicate that the current practice of using LLM-generated code is typically limited to either demonstrating high-level concepts or providing examples in documentation, rather than to be used as production-ready code. These findings indicate that there is much future work needed to improve LLMs in code generation before they can be integral parts of modern software development.
\end{abstract}

\begin{document}

\maketitle

\vspace{-5pt}

\section{Introduction}

The field of Artificial Intelligence (AI) has seen a major paradigm shift, characterized by powerful Large Language Models (LLMs) ~\cite{floridi2020gpt}. Recently, LLMs such as CodeGPT \cite{lu2021codexglue}, CodeParrot \cite{tunstall-2022}, and Codex \cite{chen2021evaluating} have
demonstrated their ability to facilitate code completion~\cite{xu2022systematic}, source code mapping~\cite{li2023starcoder}, system maintenance,~\cite{wang2023software} and other related Software Engineering tasks. Their contribution to software development is further reinforced by the 
iterative improvement through the collaboration between humans and AI~\cite{shin-no-date}.


However, amidst the promising developments, it is not clear how practical the integration of these LLMs into real-world production software development is, as prior work only demonstrates LLMs' potential in research settings. Specifically, the use of ChatGPT has not been studied as well as other code models yet \cite{Copilot}. Hence, it is imperative to critically evaluate the limitations and practicality of applying ChatGPT in real-world data practice.

To address this evaluation gap,
this paper focuses on two key research questions (RQs):
\begin{itemize}[leftmargin=*]
    \item RQ1: How do the developers interact with ChatGPT for code generation?
    \item RQ2: How helpful is the code generated by ChatGPT in assisting developers?
\end{itemize}
RQ1 seeks to understand the structure of conversations between developers and ChatGPT during code generation which indicates 
the dynamics of collaboration and can affect the quality of code produced. It will also help identify future direction on how to better interact with LLMs through prompts, in this domain.
RQ2 explores deeper into the usage of the generated code and studies to what extent the output of ChatGPT is useful for developers.

The main contributions of this work are:
\begin{itemize}[leftmargin=*]
    \item A detailed analysis of the real-world interactions between developers and ChatGPT
    \item An empirical examination of the usage of generated code resulted from developers and ChatGPT
    \item A publicly available dataset of developers and ChatGPT interactions, labelled with the prompt types and the final use cases of the generated code.

\end{itemize}

\vspace{-5pt}

\section{Study design}
\label{sec:approach}

The main goal of this study is to gain valuable insights into how ChatGPT are now being utilized as a tool in real-world code practice, which can help direct future research in this domain. 
We focus on answering two key Research Questions (RQs). RQ1 seeks to investigate the characteristics of the interaction between the developers and LLMs.
Specifically, in RQ1.1, we explore where exactly in the development process ChatGPT is being used. RQ1.2 studies the length of conversation between the developers and ChatGPT as the average number of prompt-response rounds needed to reach a satisfactory conclusion.
And finally, RQ1.3 is designed to identify the common categories in which developers improve ChatGPT's output using different prompting strategies.

RQ2 targets to understand how and to what extent developers utilize generated code in production.

\vspace{-5pt}
\subsection{DevGPT Dataset}
To answer the mentioned RQs, 
we utilize the DevGPT~\cite{naist-se-no-date} dataset. It is formed by searching with the keyword ``https:\\//chat.openai.com/share/'' in the GitHub GraphQL API to identify mentions of shared links sourced from software development artifacts, such as source code, commits, issues, pull requests, discussions, and Hacker News threads. Eventually, DevGPT comprised 17,913 prompts and ChatGPT's responses, which encompass 11,751 code snippets. The dataset provides six snapshots at six specific points from July 24th to August 31st, 2023.

\vspace{-5pt}
\subsection{RQ1 Design}

To answer RQ1, we followed two steps: (a) Automated data cleaning and (b) Manual labelling.


\smallskip \noindent \textbf{\textit{Automated data cleaning: }}
The original DevGPT dataset comprised ChatGPT Sharing URLs from GitHub and Hacker News references. Since this study focuses
on interactions between developers and ChatGPT on GitHub, we
filter out conversations from Hacker News sources. 
DevGPT includes ChatGPT conversations 
linked from five GitHub artifacts: 
source \textbf{Code files}, \textbf{Commits}, \textbf{Issues}, \textbf{Pull requests}, and \textbf{Discussions}. The DevGPT dataset contains JSON files documenting the details of each ChatGPT interaction. Parsing the ``Conversations'' attribute in the JSON files allows us to 
extract the total number of conversation rounds for each ChatGPT 
interaction. However, during our analysis, we identified instances where conversations included irrelevant prompt-response rounds such as expressions of gratitude and greetings to ChatGPT. For instance, following a successful solution, developers frequently express their appreciation by stating, ``Thank you, the approach worked'', bringing the conversation to a close. Alternatively, they might commence the interaction with a greeting such as, ``Hi, can you help me with the problem?''. To eliminate such prompts, two co-authors conducted manual reviews of the conversations and excluded a total of 344 irrelevant conversation rounds from the 2,299 conversations related to code generation. To ensure the conversations were correctly reviewed, each conversation was reviewed by both authors and any disagreement was resolved with a discussion.

\smallskip \noindent \textbf{\textit{Manual labeling:}}
To address RQ1.3, we looked into how developers use the prompts to have a conversation with ChatGPT for code generation and how they evolve their prompts to improve the outputs. We used an existing list of categories from a recent study \cite{shin-no-date} where the authors have looked into the same problem but within a small interview study. 

In our case, due to the volume of data to label, we utilize the crowd-sourcing process to label each data sample (each conversation) using one of the labels~\cite{shin-no-date} as shown in Table~\ref{tab:code_cat}.

To run the crowd-sourcing, first, we developed a description of the task with clear instructions and examples based on the definitions above. The task is assigning each pair of generated code (accessed via the given ChatGPT sharing link) and changed source code (access via the given source Github link) to one of the 9 categories as shown in Table~\ref{tab:code_cat}.
Using crowd-sourcing methods, we distributed the 2,299 valid data entries to volunteers whom we recruited from our local software engineering community.
The crowdsourcing model leverages the experience and diversity of our local software engineering community to ensure the high-quality and professional classification of data.
Each pair of data was reviewed by three developers and the majority was used as the label. If all three initial judgments were completely different, the opinions of another two developers were added. 
We continue adding two more opinions until a category gets at least two votes more than any other category.


\vspace{-5pt}
\subsection{RQ2 Design}

To answer RQ2 we look into code generation conversations that are shared in Pull Requests. 
As shown in ~\hyperref[tab:CG]{Table 1}, there are 189 distinct conversations pertaining to code generation within the Pull Request category. To efficiently determine whether the generated code is implemented in the actual production system, we utilized the ``View review changes'' functionality for each merged Pull Request on GitHub. Consequently, we excluded 51 Pull Request records that were not merged into the primary project, resulting in 138 merged Pull Requests forming the basis for our data in terms of usage classification.

During our initial investigation, we found that
some of the generated code snippets are used in the master branch (either directly as a copy, with some modifications, or integrated only into the instructional document), while a lot of them were not used at all. 
Thus we assign each generated code snippet with one of the following labels, manually (leveraging the View review changes functionality of GitHub) labeled by two co-authors (independently, and then double checked and discussed if disagreed until resolved):
\begin{itemize}[leftmargin=*]
    \item \textbf{Exact Match} denotes instances where developers employed the code exactly as generated by LLM without any modifications.
    \item \textbf{Modified Code} signifies cases where developers made slight alterations to certain sections of the generated code to align with the production code.
    \item \textbf{Document} encompasses situations where developers incorporated the generated code into instructional documents, such as README or test case files.
    \item \textbf{Supplementary Info} characterizes scenarios where developers did not directly employ the generated code but provided it as supplementary information or as a source of conceptual inspiration. 
\end{itemize}
To verify the accuracy of the labels, each code snippet was reviewed by both authors and any disagreement was resolved with a discussion.

\vspace{-5pt}
\section{Result and Discussion}

In this section, we will present our findings on RQ1 and 2. 

\vspace{-5pt}
\subsection{RQ1: How do the developers interact with ChatGPT for code generation?}
\label{sec:RQ1}

\begin{table}
    \centering
    \vspace{-5pt}
    \caption{The number of analyzed conversations between ChatGPT and developers.
    }
    \label{tab:CG}
    \vspace{-5pt}
    \resizebox{0.8\columnwidth}{!}{%
    \begin{tabular}{l r r r r}
    \toprule
    GitHub categories & CG & Non-CG & Both & \% of CG \\
    \midrule
    Pull Request & 189 & 79 & 268 & 70.5\% \\ 
    Issue & 362 & 154 & 516 & 70.2\% \\
    Discussion & 36 & 23 & 59 & 61.0\% \\
    Commit & 660 & 10 & 670 & \textbf{98.5\%} \\
    Code File & 1052 & 958 & \textbf{2010} & 52.3\% \\
    \midrule
    Total & 2299 & 1224 & 3523 & 65.3\% \\
    \bottomrule
    \end{tabular}
    }
\end{table}

To investigate the interaction between developers and ChatGPT across different timestamps, we extracted and filtered all accessible and distinct conversations from each of the six snapshots in the DevGPT dataset. Parsing JSON files and choosing conversations with multiple instances of ``ListofCode'' in ChatGPT's responses, we categorize a conversation as related to code generation (CG) if it contains at least one code snippet in ChatGPT's responses. Table~\ref{tab:CG} displays the number of conversations categorized as related to code generation (CG) and not related to code generation (Non-CG). The conversations are split into five GitHub categories (five rows) with a total number of analyzed conversations listed in the Total row. The five categories correspond to the sources of conversations (i.e., \textbf{Pull Request}, \textbf{Issue}, \textbf{Discussion}, \textbf{Commit}, and Source \textbf{Code file}. For each category, the number of CG, Non-CG, and both types of conversations along with the percentage of CG-related conversations are listed in the corresponding row.

\smallskip\noindent \textbf{RQ1.1: What is the distribution of conversations from different GitHub sources?}
Overall, we analyze 3523 conversations with 65.3\% of them are CG-related. Our analysis shows that while, the majority of conversations occurred in the context of code files (2010), only about half (52.3\%) of them are directly associated with code generation which is the lowest among all categories. In contrast, the Commit category exhibits a markedly high association with code generation 
with the majority of conversations (98.5\%) involving some form of code snippet generation.
Our manual analysis shows that, in code file conversations, developers often seek clarification on certain code snippets rather than asking for code generation. On the other hand, during Commit-related discussions, developers frequently ask ChatGPT to improve provided code snippets. A further explanation is mentioned in RQ1.2.





\noindent \textbf{RQ1.2: How many prompt-response rounds does it take on average upon conclusion?}
To further analyze the characteristics of the conversations between developers and ChatGPT in a code-generating context, we investigate the number of prompt-response rounds required for each CG-related conversation to reach its conclusion. Due to the presence of irrelevant conversation rounds at times, we manually analyze each CG-related conversation and count the number of rounds required for the code-generation question to be satisfactorily answered by ChatGPT. 
Figure~\ref{fig:Conversation_Round_Distributions} shows the box plot that represents the distribution of the number of required prompt-response rounds for the five GitHub categories. The means are represented by the red stars and the medians are represented by the orange lines.

Figure~\ref{fig:Conversation_Round_Distributions} highlights that conversations in the Commit context lasted on average only 2.4 rounds while conversations in the Code file context lasted on average significantly longer at 10.4 rounds. 


\begin{table}[]
\vspace{-5pt}
\caption{The number of studied code generation conversations per prompt category, as defined by Shin at el\cite{shin-no-date}
}
\label{tab:code_cat}
\vspace{-5pt}
\resizebox{0.65\columnwidth}{!}{%
\begin{tabular}{lr}
\toprule
\textbf{Categories}               & \textbf{Count} \\
\midrule
Request improvements              & 613            \\
Request more description          & 353            \\
Add specific instructions         & 351            \\
Ask questions to find correct way & 297            \\
Add more context                  & 256            \\
Request examples                  & 155            \\
Request verification              & 93            \\
Point mistake then request fix    & 112            \\
Request another generation        & 69             \\
\midrule
\textbf{Total}                    & 2299           \\
\bottomrule
\end{tabular}
}
\vspace{-5pt}
\end{table}

\smallskip\noindent\textbf{RQ1.3: How do developers use ChatGPT to improve their code?}
As described in Section~\ref{sec:approach}, to investigate how developers use ChatGPT to improve their code, we analyze 
the six checkpoints from DevGPT. We manually categorized the data based on the nine conversational prompting categories, as proposed by Shin et al.~\cite{shin-no-date}. Table~\ref{tab:code_cat} shows the number of studied conversations per category. 

Most of the conversations (more than a quarter) are about ``\textbf{Request improvements}''
while only a few are about ``\textbf{Request examples}'' or ``\textbf{Request verification}''. After consolidating the findings from RQ1.2, it was observed that 64.98\% of conversations sourced from Commit focused on \textit{Code Improvement}, including ``request improvements'', ``Add specific instructions'', ``Add more context'', ``Request verification'', ``Point mistake then request fix'' and ``Request another generation''. Additionally, 35.02\% of conversations within Code files centred around \textit{Code Clarification and Explanation}, such as ``Request more description'', ``Ask questions to find correct way'' and ``request examples''. The result explains that, when developers provide specific code snippets directly to ChatGPT for improvement (in the Commit context), the desirable outcomes are likely achieved quicker with fewer clarifications or explanations of the prompts. Conversely, more prompt-response rounds are needed to explain and clarify code functions in Code file conversations.
Table~\ref{tab:code_cat} shows that
``\textbf{Request another generation}'' conversation is rare. We suspect that since GPT-3.5 is based on transformer architecture, it will tend to ``remember'' recent conversations and ``forget'' older context\cite{gu2020improving}. This leads to additional generation requests to confuse ChatGPT into generating bad-quality code, hence the developers avoided this option.



\subsection{RQ2: How helpful is the code generated by ChatGPT in assisting developers?}

\begin{figure}
  \vspace{-5pt}
  \includegraphics[width=0.9\linewidth]{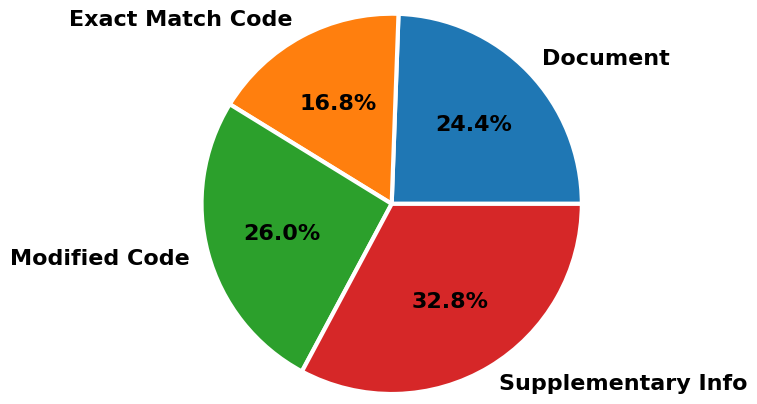}
  \vspace{-5pt}
  \caption{The proportion of different generated code usages in real-world projects
  }
  \label{fig:Code_used}
  \vspace{-5pt}
\end{figure}

As described in Section~\ref{sec:approach}, we categorize the conversation based on how the generated code snippets are used (if at all) in the master branch of the corresponding project. Figure~\ref{fig:Code_used} shows the proportions of conversation in each usage category.

As shown in Figure~\ref{fig:Code_used}, in most cases, the generated code is \textbf{Supplementary Info} to developers, perhaps due to the inferior quality. For instance, a commit conversation indicates that the generated code is considered unhelpful, causing a notable slowdown instead of enhancing code execution speed. Sometimes (24.4\%), the code snippets are used in instructional \textbf{Documents} or test cases. For instance, developers could request ChatGPT to generate a README file by providing a prompt that includes descriptions of the GitHub repository.


In 16.8\% of conversations, the generated code snippets are \textbf{Exact Matches} to the code in the master branch of the corresponding projects. In such cases, the snippets are used exactly or with very trivial modifications. For instance, developers submit a prompt that encompasses a segment of the source code along with details about an exception they encountered. They request ChatGPT to identify and modify the source code. The generated code is subsequently employed directly in the source code without any further modifications. 



In some other cases (26\%), the generated code snippets are heavily \textbf{Modified} before being included in the source code of the corresponding projects. For instance, developers might rename variables or add exception handling before merging into the primary project.


  

\subsection{Discussion}
Survivorship bias poses a validity threat in the DevGPT conversations dataset, as it is constructed from retained conversations, potentially overlooking a broader spectrum of developer interactions. The filtering bias, where developers are less likely to include ChatGPT URLs in commit messages if the generated code’s quality is too low. This may skew the dataset towards resolved discussions. Therefore, conclusions like "Commits often require fewer prompt-response rounds on average" and "a majority of conversations occurred in the context of code files," may be based on incomplete information. We acknowledge dataset limitations based on empirical experience during manual dataset processing, but lack of quantitative evidence. Future work might need to address this threat is essential for more robust conclusions.

  


\balance
\vspace{-5pt}
\section{Related work}

\smallskip \noindent \textbf{Evaluation of LLMs in generating code: }
Recent improvements ~\cite{zhong2023study, vaithilingam2022expectation} in LLMs'
code generation ability
has brought 
attention to evaluating LLM-generated code from research communities.
LLMs have been evaluated in terms of generated code correctness~\cite{liu2023your,austin2021program,chen2021evaluating}, robustness~\cite{zhong2023study}.
Prior work~\cite{zhong2023study,feng2023investigating} has also conducted human studies to evaluate how code generation supports developers in their tasks. These studies have demonstrated the potential of LLMs in code generation, however, they were all conducted in research settings. Hence the LLMs' effectiveness in real-world scenarios is inconclusive.~\cite{sridhara2023chatgpt}. This study focuses on accessing the code generation capabilities of LLMs using data scraped from an open-source platform dedicated to software development.

\smallskip \noindent \textbf{Interaction with LLMs regarding source code: }
Prior studies analyze the interaction of developers with LLMs to improve the usability~\cite{barke2023grounded} and reliability~\cite{ross2023programmer} of code generation. Prior studies discovered that LLMs reduce distractions during software development~\cite{liang2023understanding} and introduce fewer vulnerabilities~\cite{sandoval2022security}. However, studies have shown that both novices (e.g., students)~\cite{prather2023s} and experienced developers~\cite{sarkar2022like} can struggle to use LLMs on programming tasks. Our study provides some insight on all of the above aspects when LLMs are used in real-world development settings.


\smallskip \noindent \textbf{Code generation on software development platforms: }
StackOverflow and GitHub are the two most popular platforms used in software development for purposes such as communication, question-answering, and collaborative efforts~\cite{wang2013empirical}. Recently, LLMs have been used to generate code addressing problems raised on these platforms.~\cite{zhong2023study,lee2023github}. However, there have been limited prior studies that analyze the actual usage of such generated code in real-world workflow\cite{77bdac06448940a9a2ca4024277b17fa}. This study aims to bridge that gap and provide additional insight into how LLM code generation is integrated into software development.

\vspace{-5pt}
\section{Conclusion}
Large language models such as ChatGPT have shown promises in code generation.

Our study reveals that while a majority of ChatGPT conversations within code files focus on discussions, only half are related to code generation. Commit-related interactions predominantly revolve around code improvement, with fewer prompt-response rounds. Developers primarily use ChatGPT for requesting improvements and tend to avoid additional code generation within the same conversation to prevent confusion. Notably, 32.8\% of generated code is not used, emphasizing the need for further exploration of the practical utility of AI-generated code. These insights provide valuable considerations for refining AI-assisted development tools and enhancing collaboration between developers and AI systems. Much improvement is needed before LLMs become an integral part of modern software development.

\vspace{-5pt}
\section{Acknowledgement}
This work was partly supported by NSERC and Alberta Innovates.


\newpage

\balance

\bibliographystyle{ACM-Reference-Format}
\bibliography{refs}

\end{document}